\documentclass[
 reprint,
 superscriptaddress,
showpacs,preprintnumbers,
 amsmath,amssymb,
 aps,
 pra
]{revtex4-1}

\usepackage{amsmath,amssymb,bbm,epsfig}
\usepackage{enumerate}

\usepackage{color}
\usepackage[usenames,dvipsnames,svgnames,table]{xcolor}
\usepackage[colorlinks=true,
            linkcolor=blue,
            urlcolor=blue,
            citecolor=blue]{hyperref}

\newcommand{\be}{\begin{equation}}
\newcommand{\ee}{\end{equation}}
\newcommand{\ba}{\begin{array}}
\newcommand{\ea}{\end{array}}
\newcommand{\bqa}{\begin{eqnarray}}
\newcommand{\eqa}{\end{eqnarray}}

\newcommand{\vc}{\boldsymbol{c}}

\newcommand{\vd}{{\dagger}}
\newcommand{\va}{\boldsymbol{c}}

\begin{document}

\title{Structure-dynamics relation in shaken optical lattices}

\author{Albert Verdeny}
\author{Florian Mintert}

\affiliation{Department of Physics, Imperial College London, London SW7 2AZ, United Kingdom}
\affiliation{Freiburg Institute for Advanced Studies, Albert-Ludwigs-Universit\"at, Albertstrasse 19, 79104 Freiburg, Germany}

\begin{abstract}
Shaken optical lattices permit to coherently modify the tunneling of particles in a controllable manner.
We introduce a general relation between the geometry of shaken lattices and their admissible effective dynamics. Using three different examples, we illustrate the symmetries of the emerging tunneling rates. 
The results provide a clear framework to understand the relation between lattice geometry and accessible dynamics, and a tool to straightforwardly derive truncated effective Hamiltonians on arbitrary geometries.
\end{abstract}

\maketitle
\hypersetup{%
    pdfborder = {0 0 0}
}

\section{Introduction}

Advances in the control of quantum systems \cite{Bloch12} have permitted to move the scope of experiments from corroborating fundamental effects to, most recently, designing systems that target specific effects such as geometric frustration \cite{Kim10,Nixon13} or topological energy bands \cite{Atala13,Jotzu14,Aidelsburger15,Duca15}. 

A common approach to modify the properties of a system is by means of external periodic driving \cite{Goldman14,Bukov14ar}. In a suitable fast-driving regime, driven systems can be used to mimic the dynamics of time-independent systems with properties determined by the amplitude and frequency of the driving \cite{Jaksch98,Sorensen99,Lignier07,Aidelsburger13,Miyake13} or, more generally, its temporal shape \cite{Struck12,Verdeny14}.

In the domain of many-body physics, shaken optical lattices offer a particularly tunable platform for quantum simulations. Pioneer experiments in this context focused on the effects of the lowest order term in a high-frequency expansion of the effective Hamiltonan, and led to the observation of coherent suppression of tunneling \cite{Lignier07} and synthetic magnetism \cite{Struck11,Struck12}. Most recently, the interplay of the leading and higher order terms has been exploited in order to simulate tunneling processes that lead to topological energy bands \cite{Jotzu14}. Taking advantage of such higher order terms broadens the possibilities for quantum simulations substantially, since the range of available physical processes that contribute to the effective Hamiltonian grows with this order.

The increasing possibilities, however, also come with increasing requirements towards the driving that needs to be chosen such that desired processes enter the effective Hamiltonian with substantial weight whereas undesired processes are suppressed.
This can be done with an appropriately chosen temporal shape of the driving \cite{Verdeny15}, but the underlying geometry of a lattice provides limitations, and also potential for such control.

In this article we investigate the influence of the lattice geometry on the existence or absence of higher order terms in the effective Hamiltonian. Specifically, we discuss how the symmetries of the lattice geometry lead to strict selection rules resulting from destructive interference.
These rules indicate which lattice geometries exclude the realization of certain processes and how undesired processes can be avoided completely.
In Secs. \ref{sec:effham} and \ref{sec:model}, we briefly review the notion of effective Hamiltonians and the model of shaken lattices, respectively. The main results on the geometry dependence of the effective dynamics are presented in  Sec. \ref{sec:geometry}.
Finally, in Sec. \ref{sec:examples} we exemplify the symmetries of the emergent tunneling rates with three different geometries: a zig-zag chain, a kagome lattice and a Lieb lattice.

\section{Effective Hamiltonians}\label{sec:effham}

Effective Hamiltonians of driven systems naturally arise as consequence of Floquet theorem \cite{Floquet83}, which asserts that the time-evolution operator of a periodic Hamiltonian $H(t)=H(t+T)$ can be expressed as
\begin{eqnarray}\label{Floquet}
U(t)=U_F^\dagger(t)e^{-iH_{\rm eff}t}U_F(0),
\end{eqnarray}
with a periodic unitary $U_F(t)=U_F(t+T)$ and an Hermitian time-independent operator $H_{\rm eff}$. This operator defines the effective Hamiltonian of the system in the gauge $U_F(0)= e^{-iF}$.  
In a suitable fast driving regime, where the driving frequency $\omega=2\pi/T$ is the largest energy scale of the system, $U_F^\dagger(t)$ is nearly constant ($U_F^\dagger(t)\approx U_F^\dagger(0)$) with fast but small fluctuations.
Any $F$ satisfying $\lim_{w\to\infty}F=0$ ensures that $U_F^\dagger(t)$ approximately equals that identity, such that $e^{-iH_{\rm eff}t}$ provides a good approximation of $U(t)$.

The representation in Eq. (\ref{Floquet}) can be found perturbatively in powers of $\omega^{-1}$ using different methods  \cite{Magnus54,Rahav03,Verdeny13,Goldman14}. The effective Hamiltonian may then be expressed as
\begin{equation}\label{Hefftrunc}
H_{\rm eff}=H_{\rm eff}^{(0)}+H_{\rm eff}^{(1)}+H_{\rm eff}^{(2)}+\dots,
\end{equation}
with the superscript $^{(k)}$ indicating the order $O(\omega^{-k})$ of each term. 
Similarly, the generator of the gauge can  be written as 
\begin{eqnarray}
F=F^{(1)}+F^{(2)}+F^{(3)}\cdots,
\end{eqnarray}
where the expansion contains no zeroth order term $F^{(0)}$ due to the requirement that $U_F$ remains close to the identity. If a driven Hamiltonian leads to a zeroth order contribution $F^{(0)}$, as in the case of strongly shaken lattices that we shall consider, one can work in a moving frame defined by a unitary $U_I(t)$ where no such complication arises.

The two leading-order terms of the effective Hamiltonian then read $H_{\rm eff}^{(0)}=H_0$ and
\begin{eqnarray} 
\begin{split}
H_{\rm eff}^{(1)}&= \dfrac{1}{\omega}\sum_{n=1}^\infty \dfrac{1}{n} \left( [H_n,H_{-n}]+[H_{-n}-H_n,H_0]\right) \\\label{h1}
&+i[F^{(1)},H_0],
\end{split}
\end{eqnarray}
with the Fourier components $H_n=(1/T)\int_0^T H(t) e^{-in\omega t} dt$ and the lowest-order term of the gauge generator $F^{(1)}$. 
The choice of gauge enters in the first-order expression of the effective Hamiltonian but it does not affect its eigenvalues up to the order of expansion. Typical conventions are the Floquet gauge $F=0$, or the choice \cite{Goldman14}
\begin{equation}\label{gauge}
F^{(1)}=\sum_n\dfrac{i}{\omega n}(H_{-n}-H_{n}),
\end{equation}
which leads to an effective Hamiltonian $H_{\rm eff}^{(1)}$ with no dependence on the static Fourier component.

\section{The model}\label{sec:model}
We consider non-interacting particles on a periodically driven lattice described by the Hamiltonian
\begin{eqnarray}\label{Hdriven}
\tilde{H}(t)=H_s+H_d(t).
\end{eqnarray}
The system Hamiltonian $H_s$ characterizes the tunneling of the particles on a lattice with $d$ sites in each unit cell ($d$-point basis). In a single-band tight-binding description, it reads \cite{Jaksch98}
\begin{eqnarray}\label{tildeH}
H_s=\sum_{ij}  \va_i^\vd  J_{ij} \va_j,
\end{eqnarray}
in terms of the $d\times d$ tunneling matrices $J_{ij}$ and the vector creation and annihilation operators 
\begin{eqnarray} \label{vectorCA}
\begin{split}
\va_i^\vd &=(c_{i,s_1}^\dagger,\cdots,c_{i,s_d}^\dagger),\\
\va_i &=(c_{i,s_1},\cdots,c_{i,s_d})^T.
\end{split}
\end{eqnarray}
The indices $i$ and $j$ denote sites of the underlying Bravais lattice and the pair of indices $\{i,s_k\}$ characterize the physical lattice site $s_k$ of the $i$th unit cell.
The matrix elements $J_{ij}|_{s_p s_q}$ thus indicate the tunneling amplitudes among physical sites.
 Infinite lattice size or periodic boundary conditions are assumed.
The lattice site creation and annihilation operators $c_{i,s_k}^{(\dagger)}$ satisfy the usual (anti-)commutation relations $[c_{i,s_p},c^\dagger_{j,s_q}]_{\pm}=\delta_{ij,s_p s_q}$,  where the commutator applies to bosons and the anticommutator to fermions.

The time-dependent part of the Hamiltonian $H_d(t)$ describes the periodic driving of the system and is given by \cite{Dunlap86}
\begin{eqnarray}
H_d(t)=\sum_i \va_i^\vd E_i(t)  \va_i
\end{eqnarray}
with the diagonal matrix
\begin{eqnarray}
E_i(t)=\mbox{diag}(-\textbf{r}_{i,s_1}\cdot \textbf{F}(t),\cdots,-\textbf{r}_{i,s_d}\cdot \textbf{F}(t) )
\end{eqnarray}
expressed in terms of the external driving force $\textbf{F}(t)$ and the lattice site positions $\textbf{r}_{i,s_k}$. 
We focus on a strong driving regime where the amplitude of the driving force is comparable to the driving frequency. 
In order to confine the magnitude of the time-dependent oscillations it is convenient to perform a unitary transformation \cite{Eckardt09} and work in the frame defined by
\begin{eqnarray}\label{u}
U_I(t)=\exp\left(i \sum_i \va_i^\vd \Theta_i(t) \va_i \right),
\end{eqnarray}
where $\Theta_i(t)=\int_0^t E_i(\tau)d\tau-\langle \int_0^t E_i(\tau)d\tau\rangle_T$ and $\langle \cdot\rangle_T$ denotes time average.
The Hamiltonian in Eq.  (\ref{Hdriven}) then transforms to $H(t)=U_I\tilde{H}U_I^\dagger-iU_I\partial_t U_I^\dagger$, which we shall regard as our starting-point Hamiltonian for the remainder of the article. It reads
\begin{eqnarray}\label{Ht}
H(t)&=&\sum_{ij}  \va_i^\vd  G_{ij}(t) \va_j
\end{eqnarray}
in terms of time-dependent matrices
$G_{ij}(t)=e^{-i \Theta_i(t)} J_{ij}e^{i \Theta_j(t)}$.
Importantly, the Hamiltonian in Eq. (\ref{Ht}) has the same translational invariance as the underlying Bravais lattice of $H_s$.

\section{Geometry dependence on the effective Hamitlonian} \label{sec:geometry}


The effective Hamiltonian of the shaken lattices depends crucially on the geometry of the lattice. In this section, we will explore this geometry dependence and put special emphasis on the symmetries of the geometry-dependent tunneling processes. 

With a high-frequency expansion of the effective Hamiltonian, as introduced in Eq. (\ref{Hefftrunc}), the lowest order term corresponding to the Hamiltonian $H(t)$ in Eq. (\ref{Ht}) reads
\begin{eqnarray}
H_{\rm eff}^{(0)}=\sum_{ij}  \va_i^\vd A_{ij} \va_j,
\end{eqnarray}
where $A_{ij}=G_{ij}^{0}$ are the average or static Fourier components of the time-dependent matrices $G_{ij}(t)=\sum_n G_{ij}^n e^{in\omega t}$. Consequently, the tunneling rates of the leading order term in the effective Hamiltonian  change with respect to the undriven ones and possibly describe different physics, but no new tunneling processes appear. For instance, if the undriven system only contains nearest-neighbor (NN) tunneling, then the Hamiltonian $H_{\rm eff}^{(0)}$ necessarily contains only nearest-neighbor tunneling as well. That is, the leading order effective Hamiltonian cannot be used to engineer tunneling processes that are not present in the undriven system.

New tunneling processes that are not present in $H_s$ can nonetheless emerge in higher-order terms of the effective Hamiltonian $H_{\rm eff}^{(k)}$, $k\geq 1$.
We find that the existence of such higher order terms is closely related to the lattice geometry.
\begin{enumerate}
\item[(i)] \textbf{Bravais lattices}. For Hamiltonians defined on a Bravais lattice with a one-point basis, the effective Hamiltonian is exactly given by $H_{\rm eff}=H_{\rm eff}^{(0)}$ with all higher order terms necessarily vanishing $H_{\rm eff}^{(k)}=0$, $k\geq 1$, independently of the driving force.
\item[(ii)] \textbf{Non-Bravais lattices}. For Hamiltonians defined on a Bravais lattice with a $d$-point basis, $d>1$, higher order terms in the effective Hamiltonian expansion can potentially appear depending on the specific time-dependence of the driving.
\end{enumerate}
This result, derived in Appendix \ref{appA}, shows a straightforward distinction between those lattices geometries for which the effective Hamiltonian unavoidably has the same structure as the undriven one, and those for which the effective Hamiltonian can contain new processes and, thus, give rise to a more versatile platform for quantum simulations.

\subsection*{Symmetries of the first-order term}

Given a non-Bravais lattice, the lowest order term of the effective Hamiltonian with emergent tunneling processes is given by $H_{\rm eff}^{(1)}$ in Eq. (\ref{h1}). At this order, the particular gauge defining $H_{\rm eff}$ appears explicitly and needs to be specified. We choose it to be of the general form
\begin{equation}
F^{(1)}=\dfrac{1}{\omega} \sum_{ij} \va_i^\vd  F_{ij} \va_j
\end{equation}
where $F_{ij}$ are some tunneling matrices with the same translational symmetry as those of the undriven system, i.e. $J_{ij}$. This includes, in particular, the Floquet gauge and the one defined by Eq. (\ref{gauge}). For those gauges, the first-order term effective Hamiltonian can be written as
\begin{eqnarray}\label{heff1lat}
H_{\rm eff}^{(1)}=\sum_{ij} \va_i^\vd  B_{ij} \va_j,
\end{eqnarray}
with the tunneling matrices $B_{ij}=\sum_{l}V_{ilj}$ and
\begin{eqnarray}\label{tunmat}
\begin{split}
V_{ilj}=& \sum_{n=1}^\infty \dfrac{1}{\omega n}([G_{il}^n, G_{lj}^{-n}]+[G_{il}^{-n}-G_{il}^n, G_{lj}^{0}]) \\
+& \dfrac{i}{\omega}[F_{il},G_{lj}^{0}].
\end{split}
\end{eqnarray}
The emergent tunneling processes $B_{ij}$ can be interpreted as the sum of all possible virtual two-step processes $\mathcal{V}_{ilj}$, where a particle tunnels from the $j$th unit cell to the $i$th unit cell through an intermediate unit cell $l$.

In order to inspect fundamental properties of these processes, we consider the matrix elements $B_{ij}|_{s_p s_q}$, which determine the tunneling amplitude for a particle at a physical lattice site $j_q=\{j,s_q\}$ to tunnel to the site $i_p=\{i,s_p\}$. 
They can be written as the sum of all possible virtual two-step processes
\begin{eqnarray}\label{beta0}
B_{ij}|_{s_p s_q}=\sum_{l_r}\beta_{i_pl_rj_q},
\end{eqnarray}
where $\beta_{i_pl_rj_q}$ describes the tunneling amplitude of a particle tunneling from the site $j_q$ to the $i_p$ via an intermediate site $l_r=\{l,s_r\}$. Explicit expressions for $\beta_{i_pl_rj_q}$ can be readily derived using Eq. (\ref{tunmat}) and are given in Appendix\ \ref{appB}.

\begin{figure}[t]
  \centering
    \includegraphics[width=0.45\textwidth]{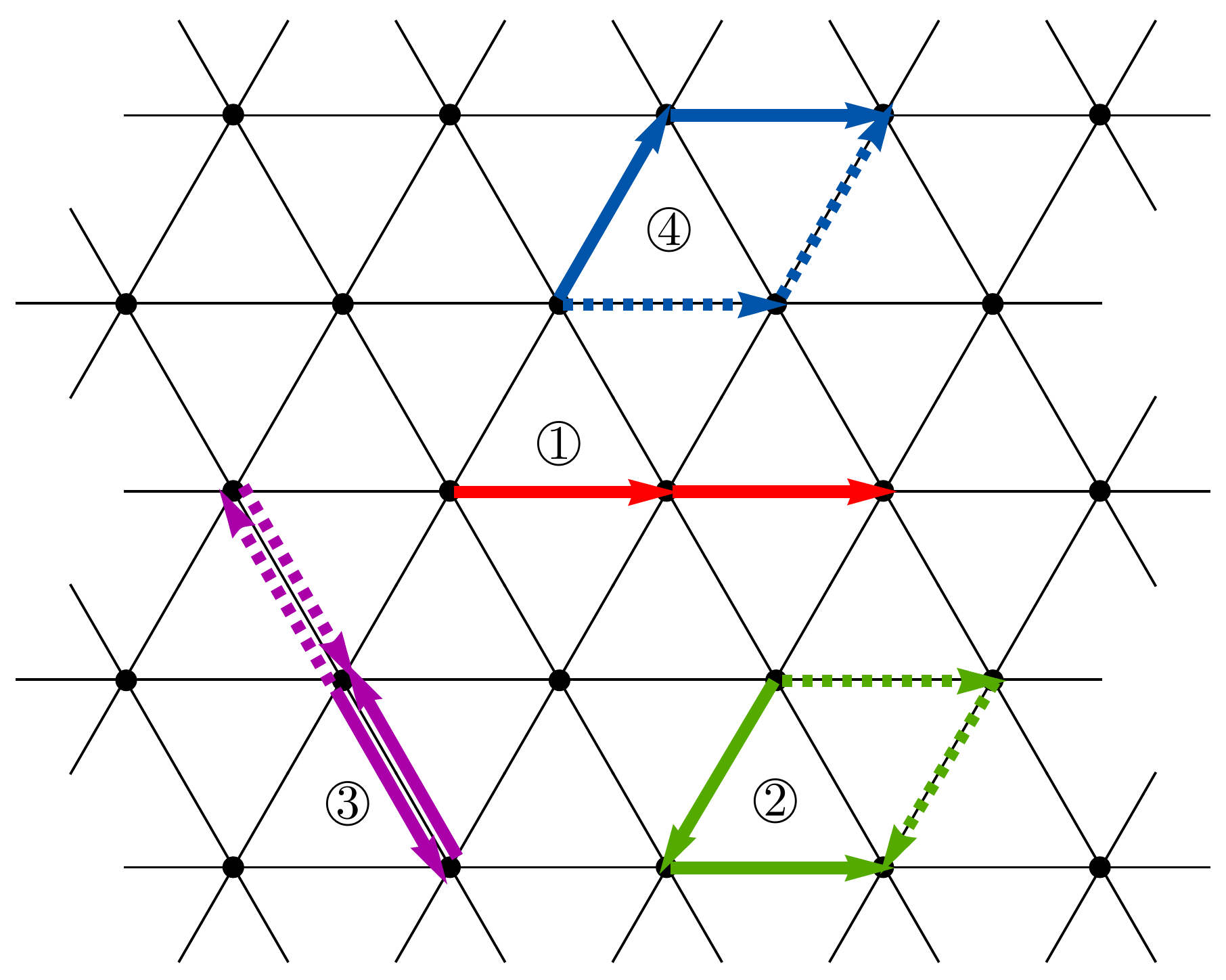}
\caption{Sketch of the triangular lattice and its two-step processes. All two-step processes can be defined either by only one vector $\{\textbf{a}_i, \textbf{a}_i\}$ (red $ \protect\raisebox{.5pt}{\textcircled{\protect\raisebox{-.9pt} {1}}}$), or by a pair of vectors $\{\textbf{a}_i, \textbf{a}_j\}$, $\textbf{a}_i\neq \textbf{a}_j$ (colors green $ \protect\raisebox{.5pt}{\textcircled{\protect\raisebox{-.9pt} {2}}}$, purple $ \protect\raisebox{.5pt}{\textcircled{\protect\raisebox{-.9pt} {3}}}$ blue $ \protect\raisebox{.5pt}{\textcircled{\protect\raisebox{-.9pt} {4}}}$) and its conjugate $\{\textbf{a}_j, \textbf{a}_i\}$ (dashed colors). For the former case, Eq. (\ref{sym}) implies that the corresponding tunneling rate vanishes. For the latter, Eq. (\ref{sym}) implies that the two tunneling process of the pair interfere destructively. This yields a vanishing first-order effective Hamiltonian $H_{\rm eff}^{(1)}=0$, in agreement with the general geometrical distinction in the beginning of Sec. \ref{sec:geometry}. This characterisitic is not specific to the triangular lattice but applies to all lattices defined on a one-point basis. In particular, it is not restricted to nearest-neighbor tunneling.}\label{Tri}
\end{figure}

An important fundamental symmetry of the quantities $\beta_{i_pl_rj_q}$ appears when the tunneling amplitudes of the undriven system $H_s$ only depend on the relative position between the sites, i.e.
\begin{eqnarray}\label{condition}
J_{ij}|_{s_p s_q}=j_{\textbf{r}_{i,s_p}-\textbf{r}_{j,s_q}}.
\end{eqnarray}
This condition is necessarily satisfied for Hamiltonians defined with a one-point basis (for which $H_{\rm eff}^{(1)}$ vanishes) and for many other lattices such as the hexagonal. Even though some non-Bravais lattices like the kagome would admit a different configuration of the tunneling rates, the condition in Eq. (\ref{condition}) is both usually considered in theoretical work \cite{Mielke92k,Zhitomirsky04,Huber10} and also implemented experimentally with optical lattices \cite{Jo12}.
When Eq. (\ref{condition}) applies, each of the virtual two-step processes can be characterized by only two vectors $\{\textbf{a}_i, \textbf{a}_j\}$ independently of the initial point. Consequently, the tunneling amplitudes $\beta_{i_pl_rj_q}$ in Eq. (\ref{beta0}) can be written as a function 
\begin{eqnarray}\label{beta1}
\beta_{i_pl_rj_q}=\beta(\textbf{a}_i,\textbf{a}_j)
\end{eqnarray}
of the two vectors  $\textbf{a}_i=\textbf{r}_{l,s_r}-\textbf{r}_{j,s_q}$ and $\textbf{a}_j=\textbf{r}_{i,s_p}-\textbf{r}_{l,s_r}$. 
The tunneling amplitudes $\beta(\textbf{a}_i,\textbf{a}_j)$ then satisfy the symmetry
\begin{eqnarray}\label{sym}
\beta(\textbf{a}_i,\textbf{a}_j)&=&-\beta(\textbf{a}_j,\textbf{a}_i),
\end{eqnarray} 
independently of the lattice geometry or external driving force, as demonstrated in Appendix \ref{appB}.
Eq. (\ref{sym}) is directly related to the fact that, for Hamiltonians defined with a one-point basis, $H_{\rm eff}^{(1)}$ vanishes, as described in Fig. \ref{Tri}. 
The symmetry in Eq. (\ref{sym}), however, also appears in Hamiltonians defined on non-Bravais lattices and leads to important constraints on the emergent tunneling processes. 
In particular, Eq. (\ref{sym}) leads to important selection rules that depend on the specific lattice geometry, as exemplified in the next section.

\section{Examples}\label{sec:examples}

Here we illustrate the symmetries introduced in Sec. \ref{sec:geometry} for the gauge defined by Eq. (\ref{gauge}) and for three different non-Bravais geometries: a zig-zag chain, a kagome lattice and a Lieb lattice. 
We consider that the underlying undriven system of all the examples contains only NN tunneling with isotropic rates satisfying the symmetry described in Eq. (\ref{condition}).

\subsection{Zig-zag chain}

\begin{figure}[t]
  \centering
    \includegraphics[width=0.45\textwidth]{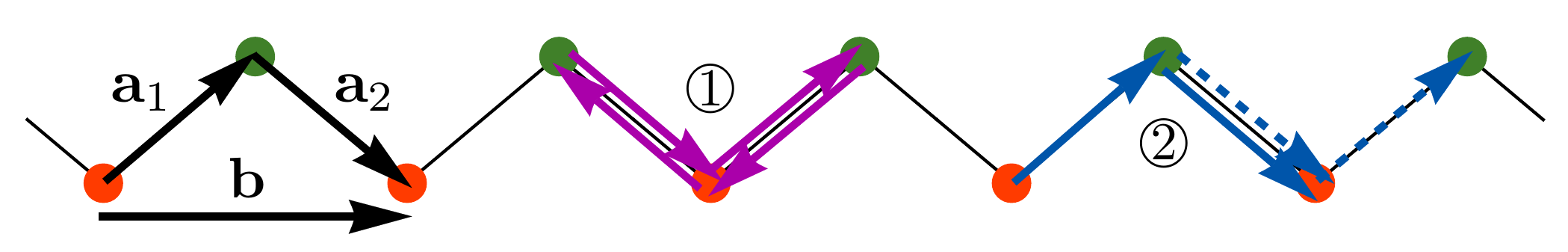}
\caption{Sketch of a zig-zag lattice and its emerging processes. The vector $\textbf{b}$ defines the primitive vector of the one-dimensional Bravais lattice. The vectors $\textbf{a}_1$ and $\textbf{a}_2$ connect neighboring sites  $s_1$ and  $s_2$, which are indicated with red and green dots, respectively. The purple $\protect\raisebox{.5pt}{\textcircled{\protect\raisebox{-.9pt} {1}}}$ arrows indicate the virtual processes giving rise to the on-site energy $b_0$.
Blue $\protect\raisebox{.5pt}{\textcircled{\protect\raisebox{-.9pt} {2}}}$ arrows illustrate two emergent next-nearest-neighbor tunneling with opposite rates.}\label{figurezz}
\end{figure}

A zig-zag chain can be defined in terms of a one-dimensional Bravais lattice with primitive vector $\textbf{b}$, and two sites per unit cell denoted by $s_1$ and $s_2$, as sketched in Fig. \ref{figurezz}. For this geometry, the Hamiltonian $H(t)$ in Eq. (\ref{Ht}) becomes 
\begin{eqnarray}\label{Hzz}
\begin{split}
H(t)&=\sum_{i}\vc^\vd_{\textbf{r}_{i}}G_{0}(t)\vc_{\textbf{r}_i} \\
&+\sum_{i}\left(\vc^\vd_{\textbf{r}_{i}+\textbf{b}}G_{1}(t)\vc_{\textbf{r}_i}+\mbox{H.c.}\right)
\end{split}
\end{eqnarray}
with the matrices
\begin{eqnarray}
G_{0}(t)=\left(
\begin{matrix}
0 &g_{-\textbf{a}_1}(t)\\
g_{\textbf{a}_1}(t)& 0
\end{matrix}
\right), \ & &  G_{1}(t)=\left(
\begin{matrix}
0&g_{\textbf{a}_2}(t)\\
0&0
\end{matrix}
\right)
\end{eqnarray}
defined through the time-dependent tunneling rates
\begin{eqnarray}\label{gt}
g_{\textbf{a}_k}(t)&=&j_0 e^{i\chi_k(t)},
\end{eqnarray}
in terms of the tunneling rates $j_0$, the time dependence $\chi_k(t)=\int_0^t d\tau \ \textbf{F}(\tau)\cdot \textbf{a}_k-\frac{1}{T} \int_0^T dt \int_0^t d\tau \ \textbf{F}(\tau)\cdot \textbf{a}_k $, and the vectors $\textbf{a}_k$ depicted Fig. \ref{figurezz}.

The lowest order term of the effective Hamiltonian $H_{\rm eff}^{(0)}$ is readily given by the temporal average of Eq. (\ref{Hzz}) and reads
\begin{eqnarray}
H_{\rm eff}^{(0)}&=&\sum_{i}\vc^\vd_{\textbf{r}_{i}}A_{0}\vc_{\textbf{r}_i} +\sum_{i}\left(\vc^\vd_{\textbf{r}_{i}+\textbf{b}}A_{1}\vc_{\textbf{r}_i}+\mbox{H.c.}\right)
\end{eqnarray}
with the tunneling matrices
\begin{eqnarray}
A_{0}=\left(
\begin{matrix}
0 &g_{-\textbf{a}_1}^0\\
g_{\textbf{a}_1}^0&0
\end{matrix}
\right),\ & \ &\ A_{1}=\left(
\begin{matrix}
0&g_{\textbf{a}_2}^0\\
0&0
\end{matrix}
\right).
\end{eqnarray}
Consequently, $H_{\rm eff}^{(0)}$ contains renormalized NN tunneling rates $g_{\textbf{a}_k}^0$ that in general depend on the direction of tunneling $\textbf{a}_k$.

The first-order effective Hamiltonian term $H_{\rm eff}^{(1)}$,  calculated from Eq. (\ref{heff1lat}), becomes
\begin{eqnarray}
H_{\rm eff}^{(1)}&=&\sum_{i} \vc^\vd_{\textbf{r}_{i}}B_0  \vc_{\textbf{r}_i}+\sum_{i}\left(\vc^\vd_{\textbf{r}_{i}+\textbf{b}}B_1 \vc_{\textbf{r}_i}+\mbox{H.c.}\right)
\end{eqnarray}
with the matrices
\begin{eqnarray}\label{Bzz}
B_0=\left(
\begin{matrix}
b_0&0\\
0&-b_0
\end{matrix}
\right),\ & \ &\ 
B_1=\left(
\begin{matrix}
b_1&0\\
0&-b_1
\end{matrix}
\right).
\end{eqnarray}
The effective on-site energies and NNN tunneling rates are given by
\begin{eqnarray}
b_0&=&\beta(\textbf{a}_1,-\textbf{a}_1)+\beta(-\textbf{a}_2,\textbf{a}_2)\\
b_1&=&\beta(\textbf{a}_1,\textbf{a}_2),
\end{eqnarray}
 in terms of
\begin{eqnarray}\label{betaexpl}
\beta(\textbf{a}_i,\textbf{a}_j)=\sum_{n=1}^\infty \dfrac{1}{n\omega} \left( g_{\textbf{a}_i}^{-n}g_{\textbf{a}_j}^{n}-g_{\textbf{a}_j}^{-n}g_{\textbf{a}_i}^{n} \right).
\end{eqnarray}
Eq. (\ref{betaexpl}) characterizes the contribution to the effective rates arising from particles following the virtual process $\{\textbf{a}_i,\textbf{a}_j\}$, i.e. tunneling from a site $\textbf{r}$ to $\textbf{r}+\textbf{a}_i+\textbf{a}_j$ through the intermediate site $\textbf{r}+\textbf{a}_i$, as described in the previous section.

The on-site energies and tunneling of the effective Hamiltonian $H_{\rm eff}^{(1)}$ have a clear interpretation.
The on-site energy $b_0$ of sites $s_1$, for instance, emerges as a consequence of the sum of the two processes $\{\textbf{a}_1,-\textbf{a}_1\}$ and $\{-\textbf{a}_2,\textbf{a}_2\}$. On the other hand, the rate $b_1$ describing the tunneling between sites $s_1$ of neighboring unit cells, results from the process $\{\textbf{a}_1,\textbf{a}_2\}$ only, as depicted in Fig. \ref{figurezz}.
Similarly, the on-site energy $-b_0$ and rate $-b_1$ in Eq. (\ref{Bzz}) result from processes with the inverse order. The different signs thus arise from the general symmetry introduced in Eq. (\ref{sym}), in agreement with Eq. (\ref{betaexpl}).
The symmetries of the emergent tunneling imply that Eq. (\ref{condition}) no longer holds for the tunneling of the effective Hamiltonian, as the rate for particles to tunnel from a site $\textbf{r}$ to $\textbf{r}+\textbf{a}_1+\textbf{a}_2$ is different to the rate for particles to tunnel from $\textbf{r}+\textbf{a}_1$ to $\textbf{r}+\textbf{a}_1+\textbf{a}_2$. 
If $\textbf{a}_1=\textbf{a}_2$, a linear Bravais lattice is recovered and Eq. (\ref{betaexpl}) implies that $b_0=b_1=0$ consistently with the geometrical distinction in the beginning of Section \ref{sec:geometry}.

\subsection{Kagome lattice}

\begin{figure}[t]
  \centering
    \includegraphics[width=0.45\textwidth]{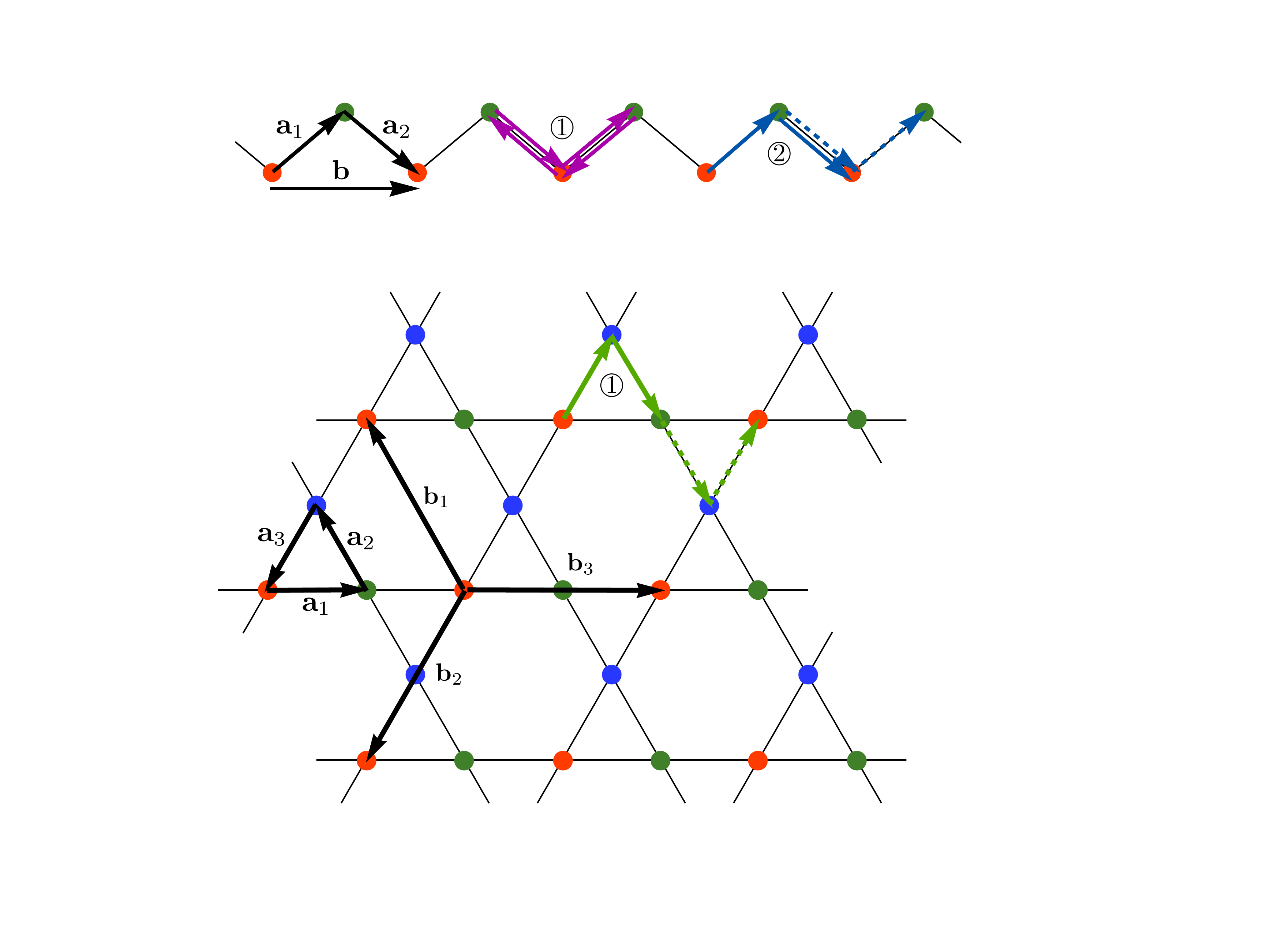}
\caption{Sketch of a kagome lattice. $\textbf{b}_1$ and $\textbf{b}_2$ denote the two primitive vectors of the underlying triangular Bravais lattice. Each unit cell contains three sites $s_1$, $s_1$ and $s_3$ in red, green and blue, respectively. The vectors $\textbf{a}_k$ connect nearest-neighbor sites. The virtual paths of two emergent nearest-neighbour tunneling with opposite rates are indicated with green $\protect\raisebox{.5pt}{\textcircled{\protect\raisebox{-.9pt} {1}}}$ arrows.}\label{figkag}
\end{figure}

The kagome lattice is one of the most studied two-dimensional non-Bravais lattices, largely due to the central role it plays in the context of geometrically frustrated physics \cite{Mielke92k,Zhitomirsky04,Huber10}. Recently, this geometry has been experimentally realized using ultra-cold atoms on an optical lattice \cite{Jo12}, which suggests that its driving could be implemented in a near future. 
The kagome lattice consists of a triangular Bravais lattice with primitive vectors $\textbf{b}_1$ and $\textbf{b}_2$, and a three-point basis, as depicted in Fig. \ref{figkag}. For this lattice, the Hamiltonian in Eq. (\ref{Ht}) becomes
\begin{eqnarray}\label{Hkl}
\begin{split}
H(t)&=\sum_{i}\vc^\vd_{\textbf{r}_{i}}G_{0}(t)\vc_{\textbf{r}_i} \\
&+\sum_{i}\sum_{k=1}^3\left(\vc^\vd_{\textbf{r}_{i}+\textbf{b}_k}G_{k}(t)\vc_{\textbf{r}_i}+\mbox{H.c.}\right)
\end{split}
\end{eqnarray}
with the vector $\textbf{b}_3=-\textbf{b}_1-\textbf{b}_2$, the matrices
\begin{eqnarray}
\begin{split}
G_0(t)&=\left(
\begin{matrix}
0&g_{\text{-}\textbf{a}_1}&g_{\textbf{a}_3}\\
g_{\textbf{a}_1}&0&g_{\text{-}\textbf{a}_2}\\
g_{\text{-}\textbf{a}_3}&g_{\textbf{a}_2}&0
\end{matrix}\right),\ 
G_1(t)=\left(
\begin{matrix}
0&0&0\\
0&0&g_{\textbf{a}_2}\\
0&0&0
\end{matrix}
\right)\\
G_2(t)&=\left(
\begin{matrix}
0&0&0\\
0&0&0\\
g_{\textbf{a}_3}&0&0
\end{matrix}
\right),\
G_3(t)=\left(
\begin{matrix}
0&g_{\textbf{a}_1}&0\\
0&0&0\\
0&0&0
\end{matrix}
\right)
\end{split}
\end{eqnarray}
and the same time-dependent tunneling rates $g_{\textbf{a}_i}(t)$ introduced in Eq. (\ref{gt}). The vectors $\textbf{a}_i$ connecting neighboring sites are represented in Fig. \ref{figkag}.

The leading effective Hamiltonian $H_{\rm eff}^{(0)}$ is readily obtained from the time average of Eq. (\ref{Hkl}) and does not contain new tunneling processes.

The first-order term $H_{\rm eff}^{(1)}$, on the other hand, is given by
\begin{eqnarray}\label{Heffkl}
\begin{split}
H_{\rm eff}^{(1)}&=\sum_{i}\vc^\vd_{\textbf{r}_{i}}B_{0}\vc_{\textbf{r}_i}\\
&+\sum_{i}\sum_{k=1}^3\left(\vc^\vd_{\textbf{r}_{i}+\textbf{b}_k}B_{k}\vc_{\textbf{r}_i}+\mbox{H.c.}\right)
.
\end{split}
\end{eqnarray}
The tunneling matrices are
\begin{eqnarray}\label{Bkl}
\begin{split}
B_0=\left(
\begin{matrix}
0&b_1^*&b_3\\
b_1&0&b_2^*\\
b_3^*&b_2&0
\end{matrix}
\right)&,\ \  \
B_{1}=\left(
\begin{matrix}
0&0&-b'^*_3 \\
b'_1&0&-b_2\\
0&0&0
\end{matrix}
\right)\\
 B_2=\left(
\begin{matrix}
0&0&0\\
-b'^*_1&0&0\\
-b_3&b'_2&0
\end{matrix}
\right)&, \  \  \
B_3=\left(
\begin{matrix}
0&-b_1&b'_3\\
0&0&0\\
0&-b'^*_2&0
\end{matrix}
\right)
\end{split}
\end{eqnarray}
defined in terms of the effective NN tunneling rates 
\begin{eqnarray}
b_1&=&\beta(-\textbf{a}_3,-\textbf{a}_2), \\
b_2&=&\beta(-\textbf{a}_1,-\textbf{a}_3), \\
b_3&=&\beta(-\textbf{a}_2,-\textbf{a}_1),
\end{eqnarray}
the NNN tunneling rates
\begin{eqnarray}
b'_1&=&\beta(-\textbf{a}_3,\textbf{a}_2),\\
b'_2&=&\beta(-\textbf{a}_1,\textbf{a}_3),\\
b'_3&=&\beta(-\textbf{a}_2,\textbf{a}_1),
\end{eqnarray}
and with $\beta(\textbf{a}_i,\textbf{a}_j)$ in Eq. (\ref{betaexpl}).
As depicted in Fig. \ref{figkag}, the relative signs in Eq. (\ref{Bkl}) are  a particular manifestation of the fundamental symmetry in Eq. (\ref{sym}) and appear due to the virtual path that particles follow.

There are two basic differences between the emergent processes of the zig-zag chain and those of the kagome lattice. On the one hand, new NN tunneling appears in the kagome lattice as a consequence of two consecutive virtual NN tunneling in a triangular geometry. On the other, no on-site energies appear because the different contributions interfere destructively. For each process $\{\textbf{a}_i,-\textbf{a}_i\}$ contributing to the on-site energy, there is a conjugate process $\{-\textbf{a}_i,\textbf{a}_i\}$. This exemplifies well how different non-Bravais lattices can induce different effective tunneling processes due to the specific details of the geometry, which opens the possibility to designing specific geometries that yield desired tunneling processes by driving them.


\subsection{Lieb lattice}

\begin{figure}[t]
  \centering
    \includegraphics[width=0.45\textwidth]{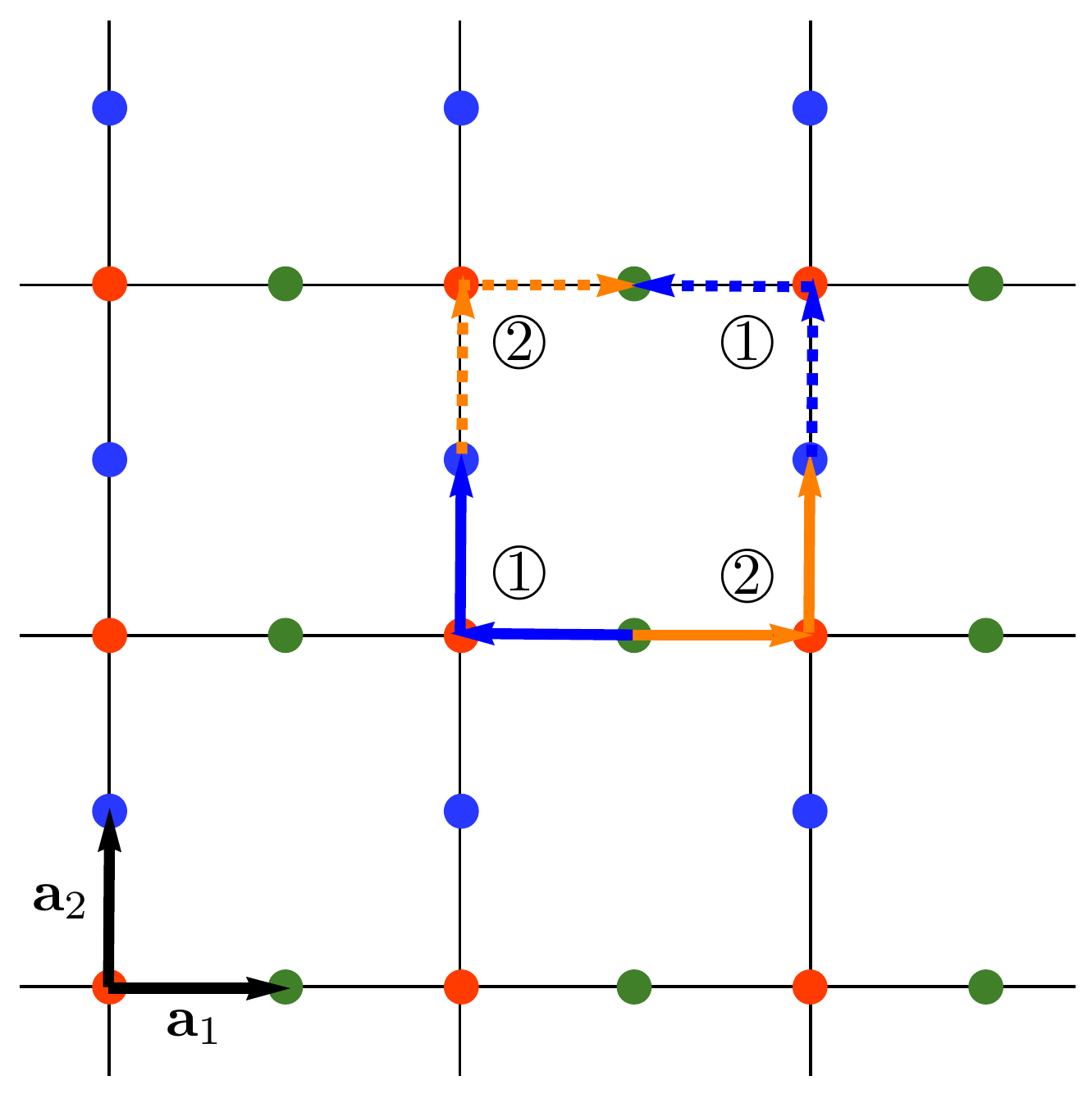}
\caption{Sketch of a Lieb lattice. The only virtual processes that can lead to a finite contribution of the emerging tunneling in the effective Hamiltonian $H_{\rm eff}^{(1)}$ correspond to all possible combinations of $\pm\textbf{a}_1$ and $\pm\textbf{a}_2$.
Virtual paths of two next-nearest-neighbor tunneling are indicated with blue $\protect\raisebox{.5pt}{\textcircled{\protect\raisebox{-.9pt} {1}}}$ and orange $\protect\raisebox{.5pt}{\textcircled{\protect\raisebox{-.9pt} {2}}}$ arrows.}\label{figlieb}
\end{figure}

With the previous examples, we have demonstrated a clear interpretation of the tunneling rates of the first-order effective Hamiltonian in terms of particles virtually tunneling through an intermediate site. In fact, this geometrical interpretation permits to infer the tunneling rates of effective Hamiltonians of shaken lattices essentially by only inspecting the geometry of the lattice, as we will describe here with particles on a shaken Lieb lattice.

The Lieb lattice is a two-dimensional face-centered square lattice, defined in terms of a square Bravais lattice and a three-point basis, as depicted in Fig. \ref{figlieb}.
Analogously to the previous sections, the driven Hamiltonian $H(t)$ in Eq. (\ref{Ht}) is defined by its time-dependent tunneling rates $g_{\textbf{a}_i}(t)$ in Eq. (\ref{gt}), characterizing the tunneling of particles from $\textbf{r}$ to $\textbf{r}+\textbf{a}_i$.

According to Eqs. (\ref{beta0}), (\ref{beta1}), and (\ref{betaexpl}), the effective tunneling rate for particles to tunnel between two arbitrary sites can be determined one by one by inspecting all the possible two-step processes connecting them and adding up all the contributions. That is the rate $b(\textbf{r},\textbf{r}+\textbf{a}_i+\textbf{a}_j)$ for particles to tunnel (or stay) from a site at $\textbf{r}$ to $\textbf{r}+\textbf{a}_i+\textbf{a}_j$ is given by
\begin{eqnarray}
b(\textbf{r},\textbf{r}+\textbf{a}_i+\textbf{a}_j)=\sum_{\{\textbf{a}_i,\textbf{a}_j\}}\beta(\textbf{a}_i,\textbf{a}_j)
\end{eqnarray}
where the sum is performed over all possible two-step processes and, in the gauge defined by Eq. (\ref{gauge}), $\beta(\textbf{a}_i,\textbf{a}_j)$ is given by Eq. (\ref{betaexpl}).
The symmetry in Eq. (\ref{sym}) essentially establishes which processes necessarily vanish and which processes can be finite for specific driving forces.
For the Lieb lattice it can be thus inferred that only NNN tunneling between green and blue sites can appear in the effective Hamiltonian $H_{\rm eff}^{(1)}$, as described in Fig. \ref{figlieb}. Moreover, the corresponding tunneling rates are $\beta(\textbf{a}_1,\textbf{a}_2)$, $\beta(\textbf{a}_2,\textbf{a}_1)$, $\beta(-\textbf{a}_1,\textbf{a}_2)$, $\beta(\textbf{a}_2,-\textbf{a}_1)$  and its complex conjugate values. All other processes such as on-site energies or NNN tunneling between translationally equivalent sites (i.e. between sites with the same color in Fig. \ref{figlieb}) do not appear because the contributions either are zero or sum up to zero, as analogously described in Fig. \ref{Tri}.

\section{Conclusions and outlook}

The structural distinction of effective Hamiltonians of shaken lattices in terms of the lattice geometry described in Sec. \ref{sec:geometry}, establishes a general framework to understand the central role that the geometry of shaken lattices plays on its effective dynamics.

The geometry dependence of the effective tunneling rates can be interpreted in terms of path-dependent virtual tunneling processes that interfere with each other. Depending on the details of the geometry, this interference can then lead to vanishing or finite effective tunneling rates. The resulting selection rules open new possibilities for quantum simulations by providing a tool to readily identify lattice geometries that contain only desired tunneling processes.

In a fast-driving regime, the lowest-order contribution to the effective Hamiltonian with geometry-dependent tunneling is given by the first-order term of a high-frequency expansion. At this order, the emerging tunneling processes result from two consecutive virtual tunneling processes of the undriven system. Consequently, if the undriven system contains only nearest-neighbor tunneling, the possible emerging processes are on-site energies and next-nearest-neighbor tunneling. The geometry dependence of the aforementioned next-nearest-neighbor tunneling can then be employed to engineer topological energy bands with non-zero Chern number \cite{Oka09,Jotzu14,He14,Verdeny15}. Our general characterization thus permits to identify a wide range of lattice geometries that can host topological phases.

Highly interesting for actual quantum simulations is the inclusion of interactions or spin-dependent tunnelling processes \cite{Jotzu14,Struck14,Jotzu15}.
The identification of structure-dynamics relations established here is rather general and not limited to the specific case of non-interacting spinless particles.
Since the approach is based on general geometric and algebraic properties
extensions to systems with different underlying dynamics may readily be obtained and also extensions to higher order processes are rather straightforward.
The possibility to identify structural properties of a quantum simulator based on its underlying geometry is thus expected to aid substantially in the design of optimal implementations thereof.

\begin{acknowledgments}
Financial support by the European Research Council within the project ODYCQUENT is gratefully acknowledged.
\end{acknowledgments}

\appendix

\section{Geometry distinction}\label{appA}

The geometrical distinction established in Section \ref{sec:geometry} is based on the commutativity of the Hamiltonian at different times and can be derived with the Magnus expansion \cite{Magnus54,Blanes09}. 

The Magnus expansion approximates the time-evolution operator as $U(t)=e^{-iM(t)}$ with 
\begin{equation}
M(t)=M_0(t)+M_1(t)+\cdots,
\end{equation}
where $M_0(t)=\int_{0}^{t} dt_1 H(t_1)$ and higher-order contributions $M_k(t)$ are given in terms of $k-1$ time integrals of $k$-fold commutators. The effective Hamiltonian of the driven system can be then obtained order by order through 
\begin{eqnarray}
H_{\rm eff}=\frac{1}{T} U_F^\dagger(0) M(T) U_F(0)
\end{eqnarray}
for any chosen gauge $U_F(0)$.

A sufficient condition for all Magnus terms $M_k(t)$, with $k\geq 1$, to vanish is that $[H(t_1),H(t_2)]=0$ \cite{Blanes09}.
Consequently, 
\begin{eqnarray}
[H(t_1),H(t_2)]=0 \implies H_{\rm eff}=H_{\rm eff}^{(0)},
\end{eqnarray}
with all higher order terms of the effective Hamiltonian exactly vanishing.

Using the (anti)commutation relations of the creation and annihilation operators of the Hamiltonian in Eq. (\ref{Ht}), the commutator of the Hamiltonian can be written as
\begin{eqnarray}
[H(t_1),H(t_2)]=\sum_{ij} \vc^\dagger_i \mathcal{M}_{ij}(t_1,t_2) \vc_j,
\end{eqnarray}
with
\begin{eqnarray}\label{M}
\mathcal{M}_{ij}(t_1,t_2)=\sum_l \left( G_{il}(t_1)G_{lj}(t_2)-G_{il}(t_2)G_{lj}(t_1)\right).
\end{eqnarray}
The translational symmetry of the Hamiltonian permits us to use the identity
\begin{eqnarray}\label{identity}
\sum_l(C_{il}D_{lj}-D_{il}C_{lj})=\sum_l[C_{il},D_{lj}],
\end{eqnarray}
which holds for translationally invariant matrices $C_{ij}$ and $D_{ij}$ as derived in Appendix \ref{appAa}. With this relation, Eq. (\ref{M}) van be rewritten in the more convenient form 
\begin{eqnarray}\label{comm}
\mathcal{M}_{ij}(t_1,t_2)&=&\sum_l [G_{il}(t_1), G_{lj}(t_2)],
\end{eqnarray}
from which the distinction between Bravais and non-Bravais lattice follows: if the Hamiltonian is defined on a one-point basis, the $G_{ij}(t)$ are scalars and the commutator necessarily vanishes. If it is defined in a $d$-point basis with $d>1$, the $G_{ij}(t)$ are matrices and the commutator can be non-vanishing. 

This distinction considers the lattice geometry through the commutativity of the tunneling matrices but it does not take into account the specific time dependence of the Hamiltonian. It is thus possible that systems within the category (ii) in Sec. \ref{sec:geometry} still satisfy $H_{\rm eff}=H_{\rm eff}^{(0)}$ due to particular driving profile.

\section{Symmetry of the tunneling rates}\label{appB}

Expressions for $\beta_{i_pl_rj_q}$ can be readily derived using Eqs. (\ref{tunmat}), (\ref{beta0}) and (\ref{identity}). They read
\begin{widetext}
\begin{eqnarray}
\begin{split}\label{explicit}
\beta_{i_pl_rj_q}&=\sum_{n=1}^\infty \dfrac{1}{n\omega}(G^n_{il}|_{s_p s_r}G^{-n}_{lj}|_{s_r s_q}-G^{-n}_{il}|_{s_p s_r}G^n_{lj}|_{s_r s_q}) \\
&+\sum_{n=1}^\infty \dfrac{1}{n\omega} ((G^{-n}_{il}-G^{n}_{il})|_{s_p s_r}G^{0}_{lj}|_{s_r s_q}-G^{0}_{il}|_{s_p s_r}(G^{-n}_{lj}-G^{n}_{lj})|_{s_r s_q}) \\
&+\dfrac{i}{\omega}(F_{il}|_{s_p s_r}G^0_{lj}|_{s_r s_q}-G^0_{il}|_{s_p s_r}F_{lj}|_{s_r s_q}).
\end{split}
\end{eqnarray}
\end{widetext}

The symmetry $\beta(\textbf{a}_i,\textbf{a}_j)=-\beta(\textbf{a}_j,\textbf{a}_i)$ introduced in Eq. (\ref{sym}) can be derived from Eqs. (\ref{explicit}) and  (\ref{condition}). The condition in Eq. (\ref{condition}) implies that the same relation applies for both $F_{ij}$ and $G^n_{ij}$, i.e.
\begin{eqnarray}
F_{ij}|_{s_p s_q}&=&f_{\textbf{r}_{i,s_p}-\textbf{r}_{j,s_q}}\label{fsym}\\
G^n_{ij}|_{s_p s_q}&=&g^n_{\ \textbf{r}_{i,s_p}-\textbf{r}_{j,s_q}}\label{gsym}
\end{eqnarray}
Eq. (\ref{fsym}) is a consequence of our choice of gauge, which has the same symmetries as $J_{ij}$.
Eq. (\ref{gsym}), on the other hand, can be derived by calculating the matrix elements of $G_{ij}(t)$ in Eq. (\ref{Ht}). 

Consider now two tranlstionally invariant matrices $C_{ij}$ and $D_{ij}$ with analogous symmetries as in Eqs. (\ref{fsym}) and (\ref{gsym}). We then define the quantity $\kappa(\textbf{a}_1,\textbf{a}_2)$ as
\begin{eqnarray}
\kappa(\textbf{a}_1,\textbf{a}_2)&=&C_{il}|_{s_p s_r} D_{lj}|_{s_r s_q}-D_{il}|_{s_p s_r}C_{lj}|_{s_r s_q}\\
&=&c_{\textbf{r}_{i s_p}-\textbf{r}_{l s_r}} d_{\textbf{r}_{l s_r}-\textbf{r}_{j s_p}}-d_{\textbf{r}_{i s_p}-\textbf{r}_{l s_r}} c_{\textbf{r}_{l s_r}-\textbf{r}_{j s_p}}\\
&=&c_{\textbf{a}_2} d_{\textbf{a}_1}-d_{\textbf{a}_2} c_{\textbf{a}_1},
\end{eqnarray}
which explicitly satisfies $\kappa(\textbf{a}_1,\textbf{a}_2)=-\kappa(\textbf{a}_2,\textbf{a}_1)$. Combining this result with Eqs. (\ref{explicit}), (\ref{fsym}) and (\ref{gsym}), the symmetry $\beta(\textbf{a}_i,\textbf{a}_j)=-\beta(\textbf{a}_j,\textbf{a}_i)$ introduced in Eq. (\ref{sym}) directly follows.

\section{Identity for translationally invariant matrices}\label{appAa}

Eq. (\ref{transmat}) can be proven by exploiting the translationally invariance of the Bravais lattice and reordering the terms of the sum. For convenience, we introduce a slightly different notation and denote translatinally invariant matrices $C_{ij}$ by $C(\textbf{r}_i-\textbf{r}_l)$, where the argument $\textbf{r}_i -\textbf{r}_l$ does not indicate an explicit dependence on the vectors but rather a label. With this notation, Eq. (\ref{identity}) can be expressed as
\begin{eqnarray}\label{transmat}
&&\sum_l (C(\textbf{r}_i -\textbf{r}_l) D(\textbf{r}_l -\textbf{r}_j)-D(\textbf{r}_i-\textbf{r}_l)C(\textbf{r}_l -\textbf{r}_j))=\nonumber\\
&&\sum_l [C(\textbf{r}_i -\textbf{r}_l),D(\textbf{r}_l -\textbf{r}_j)].
\end{eqnarray}

The translational symmetry of the Bravais lattice implies that, if the vectors $\textbf{r}_i$ and $\textbf{r}_j$ denote the position of two Bravais sites, then for each Bravais site $l$ at a position $\textbf{r}_l=\textbf{r}_i+\textbf{R}_{li}$, there exists a ``conjugate" Bravais site $l^*$ at a position $\textbf{r}_{l^*}=\textbf{r}_j-\textbf{R}_{li}$. Consequently 
$\textbf{R}_{li}=\textbf{R}_{jl^*}$.
Moreover $\textbf{r}_{l^*}=\textbf{r}_{l}$ if and only if $\textbf{R}_{li}=(\textbf{r}_j-\textbf{r}_i)/2$.

Consider a term 
\begin{eqnarray}
T_l=C(\textbf{r}_i -\textbf{r}_l) D(\textbf{r}_l -\textbf{r}_j)-D(\textbf{r}_i-\textbf{r}_l)C(\textbf{r}_l -\textbf{r}_j)
\end{eqnarray}
in the sum of Eq. (\ref{transmat}) with fixed $i$ and $j$. Then, either $\textbf{R}_{li}=(\textbf{r}_j-\textbf{r}_i)/2$ or $\textbf{R}_{li}\neq(\textbf{r}_j-\textbf{r}_i)/2$. The first case implies that the conjugate site of $l$ is itself and thus $T_l$ can directly be written as the commutator
\begin{eqnarray}
T_l=[C(\textbf{r}_i -\textbf{r}_l),D(\textbf{r}_l -\textbf{r}_j)],
\end{eqnarray}
since $\textbf{R}_{li}=\textbf{R}_{jl}$.
The second case implies that $\textbf{r}_l\neq \textbf{r}_{l^*}$ and thus there there exists another term $T_{l^*}$ in the sum such that 
\begin{eqnarray}
&&T_l+T_{l^*}=\nonumber\\
&&[C(\textbf{r}_i -\textbf{r}_l),D(\textbf{r}_l -\textbf{r}_j)]+[C(\textbf{r}_i -\textbf{r}_{l^*}),D(\textbf{r}_{l^*} -\textbf{r}_j)].
\end{eqnarray}
Therefore, the entire sum can be rewritten as
\begin{eqnarray}
\sum_l T_l=\sum_l [C(\textbf{r}_i -\textbf{r}_l),D(\textbf{r}_l -\textbf{r}_j)],
\end{eqnarray}
which completes the proof of Eq. (\ref{transmat}) and hence (\ref{identity}).

\bibliography{mybib}

\end{document}